\documentclass[10pt,conference]{IEEEtran}
\IEEEoverridecommandlockouts
\usepackage{lineno}
\usepackage{hyperref}
\usepackage{cite}
\usepackage{amsmath,amssymb,amsfonts,cases} 
\usepackage{amsmath}
\usepackage{amsthm}

\DeclareMathOperator*{\argmin}{argmin}
\usepackage{graphicx}
\usepackage{textcomp}
\usepackage{xcolor}
\usepackage{graphicx}
\usepackage{float}
\usepackage{subfigure}
\usepackage{amsmath,mleftright}
\usepackage{amsfonts,amssymb}
\usepackage{mathrsfs}
\usepackage{mathtools}
\usepackage{algorithm}
\usepackage{algorithmic}
\usepackage{bm}
\usepackage{multirow}
\usepackage{array}
\usepackage{amssymb}
\usepackage{amsmath}
\usepackage{cite}
\usepackage{url}
\usepackage{xcolor}
\usepackage{cite,graphicx,amsmath,amssymb}
\usepackage{subfigure}
\usepackage{fancyhdr}
\usepackage{mdwmath}
\usepackage{mdwtab}
\usepackage{caption}
\usepackage{amsthm}
\usepackage{setspace}
\usepackage{bm}
\usepackage{mathtools}
\usepackage{dsfont}
\usepackage{bbm}
\usepackage{framed}

\newtheorem{theorem}{Theorem}

\newtheorem{lemma}{Lemma}

\newtheorem{corollary}{Corollary}

\makeatletter
\newcommand{\biggg}{\bBigg@{3}}
\newcommand{\Biggg}{\bBigg@{3.5}}
\makeatother
\makeatletter
\allowdisplaybreaks[4]
\renewcommand{\maketag@@@}[1]{\hbox{\m@th\normalsize\normalfont#1}}%
\makeatother
\def\BibTeX{{\rm B\kern-.05em{\sc i\kern-.025em b}\kern-.08em
    T\kern-.1667em\lower.7ex\hbox{E}\kern-.125emX}}
    \expandafter\def\expandafter\normalsize\expandafter{%
    \normalsize%
    \setlength\abovedisplayskip{4pt}%
    \setlength\belowdisplayskip{4pt}%
    \setlength\abovedisplayshortskip{2pt}%
    \setlength\belowdisplayshortskip{2pt}%
}
\begin{document}
\title{Movable Antennas-Assisted Over-the-Air Computation: Dynamic and Static Design}
\author{\IEEEauthorblockN{Zhenqiao Cheng$^{\dag}$, Nanxi Li$^{\dag}$, Jianchi Zhu$^{\dag}$, Shan Yang$^{\dag}$, Chongjun~Ouyang$^{\star}$, and Xingqi~Zhang$^{\ddag}$}
$^\dag$6G Research Centre, China Telecom Beijing Research Institute, Beijing, 102209, China\\
$^{\star}$School of Electronic Engineering and Computer Science, Queen Mary University of London, London, U.K.\\
$^{\star}$Department of Electrical and Computer Engineering, University of Alberta, Edmonton AB, Canada}
\maketitle
\begin{abstract}
A novel over-the-air computation (AirComp) framework empowered by movable antennas (MAs) is proposed to significantly enhance computation accuracy. Within this framework, the joint optimization of transmit power control, antenna positioning, and receive beamforming is investigated. Two design strategies are developed: (i) a dynamic design, where MA positions are optimized based on fast-varying instantaneous channel state information (CSI); and (ii) a static design, where antenna positions are optimized using only slow-varying statistical CSI. Numerical results validate the superior MSE performance of the proposed MA-enabled AirComp framework and demonstrate its clear advantage over benchmark systems employing conventional fixed-position antennas (FPAs).
\end{abstract}
\section{Introduction}
Over the past three decades, multiple-input multiple-output (MIMO) technology has fundamentally reshaped wireless communications. In conventional multi-antenna systems, however, antenna elements are fixed in space, which restricts their ability to fully exploit spatial variations within a given transmit or receive region \cite{heath2018foundations}. This limitation becomes especially evident when the number of antennas is limited, thereby constraining spatial diversity and array gain. To overcome this drawback, recent MIMO research has entered a new phase that emphasizes reconfigurable antennas, which serve as an adaptive layer complementing traditional digital and analog beamforming \cite{heath2025tri}.

\begin{figure}[!t]
\centering
\includegraphics[height=0.12\textwidth]{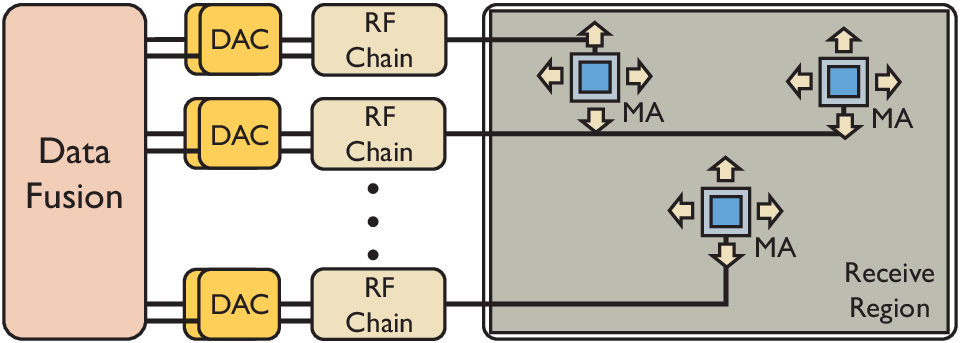}
\caption{Illustration of MA-empowered MIMO architecture.}
\label{System_Model_AirComp1}
\vspace{-20pt}
\end{figure}

One of the most representative examples of reconfigurable antennas is the movable antenna (MA) \cite{wong2020fluid,zhu2024movable}. MAs are designed to overcome the inherent limitations of conventional fixed-position antennas (FPAs) by allowing dynamic position adjustment in real time. Each MA is connected to its radio-frequency (RF) chain via a flexible cable and equipped with a position control mechanism that enables adaptive spatial relocation \cite{zhu2025tutorial}. By intelligently adjusting their positions, MAs exploit spatial diversity and effectively mitigate small-scale fading. This flexibility allows the antenna array to actively reshape the wireless channel, thereby achieving enhanced transmission performance, as illustrated in {\figurename} {\ref{System_Model_AirComp1}}.

Existing research efforts have predominantly focused on exploiting MAs to enhance communication performance \cite{zhu2025tutorial}. It is worth noting, however, that the additional spatial degrees of freedom offered by antenna position optimization can provide benefits beyond communication throughput, particularly in improving data aggregation accuracy in computation-oriented wireless networks. Motivated by this broader perspective, this work investigates the integration of MAs into over-the-air computation (AirComp) systems. In AirComp, each Internet-of-Things (IoT) sensor node (SN) collects a measurement of a parameter of interest, such as temperature or humidity, and simultaneously transmits it to a central fusion center over a shared wireless channel, where the desired function (e.g., arithmetic mean or sum) is computed directly from the superimposed signals \cite{yang2020federated}. AirComp is envisioned to play a key role in future intelligent networks, leveraging the channel's waveform superposition property to enable efficient in-channel computation for intelligence-oriented applications such as federated learning and edge inference \cite{wen2024survey}. Although several studies have explored MA-enabled AirComp systems, existing designs typically rely on perfect instantaneous channel knowledge, including both reflection coefficients and scattering angles \cite{zhang2024fluid,li2024over}. Such assumptions impose excessive channel estimation overhead and make these approaches impractical for large-scale intelligent IoT networks.

To address this research gap, this work aims to provide a comprehensive understanding of how the deployment of MAs can enhance the efficiency and accuracy of computation processes in AirComp systems. We propose an MA-empowered uplink AirComp framework that leverages the reconfigurability of MAs to optimize antenna positions, thereby achieving significant improvements in the computation mean-squared error (MSE). Specifically, we investigate two design strategies: i) a \emph{dynamic design}, where MA positions are optimized based on fast-varying instantaneous channel state information (CSI); and ii) a \emph{static design}, where the antenna positions are optimized using only slow-varying statistical CSI. For both designs, we develop efficient algorithms that jointly optimize the MA positions, digital receive beamformer, and transmit power control at each node. Numerical results demonstrate that the proposed MA-enabled designs substantially outperform conventional FPA systems in terms of computation accuracy, highlighting the potential of MAs for future intelligent AirComp networks.

\section{System Model}
\begin{figure}[!t]
\centering
\includegraphics[height=0.2\textwidth]{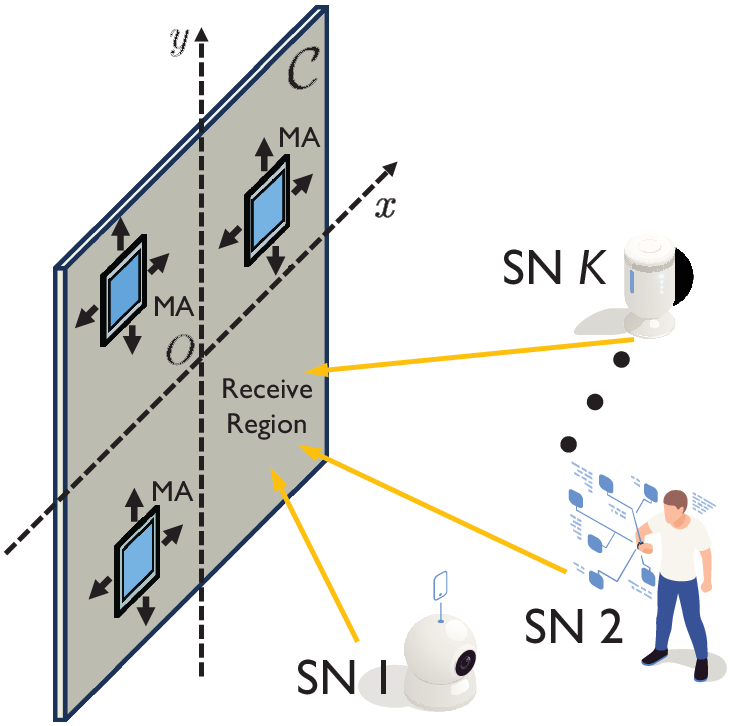}
\caption{Illustration of MA-assisted AirComp.}
\label{System_Model_AirComp}
\vspace{-20pt}
\end{figure}

We consider uplink transmission in a MA-empowered AirComp system, where $K$ IoT nodes simultaneously transmit their collected data to a fusion center for aggregation, as illustrated in {\figurename} {\ref{System_Model_AirComp}}. The fusion center is equipped with $N$ receive MAs, while each IoT node $k\in{\mathcal{K}}\triangleq\{1,\ldots,K\}$ employs a single FPA for transmission. Each MA is connected to a dedicated RF chain via flexible cables, allowing its position to be \emph{adjusted in real time} within a specified receive region, as depicted in {\figurename} {\ref{System_Model_AirComp1}}. The positions of the $n$th MA is represented by Cartesian coordinates ${\mathbf{r}}_n=[x_n, y_n]^{T}\in{\mathcal{C}}\subseteq{\mathbb{R}}^{2\times1}$ for $n\in{\mathcal{N}}\triangleq\{1,\ldots,N\}$, where $\mathcal{C}$ denotes the two-dimensional receive region within which the MAs can move freely. Without loss of generality, $\mathcal{C}$ is defined as a square region of size $A\times A$.

\subsection{Channel Model}
We assume quasi-static block-fading channels and focus on one particular fading block, within which the multipath channel components at any location in $\mathcal{C}$ remain fixed. The channel vector from SN $k$ to the fusion center is denoted by ${\mathbf{h}}_k\in{\mathbb{C}}^{N\times1}$. A field-response based channel model is considered, which can be expressed as ${\mathbf h}_{k}=[h_{k}({\mathbf{r}}_1)\ldots h_{k}({\mathbf{r}}_N)]^{{T}}$, where
\begin{subequations}\label{Channel_Model}
\begin{align}
h_{k}({\mathbf{r}})&\triangleq\sum_{\ell=1}^{L_k}\sqrt{\mu_k}\sigma_{k,\ell}{\rm{e}}^{-{\rm{j}}\frac{2\pi}{\lambda}{\mathbf{r}}^{\mathsf{T}}{\bm\rho}_{k,\ell}}\\
&=\sum_{\ell=1}^{L_k}\sqrt{\mu_k}\sigma_{k,\ell}{\rm{e}}^{-{\rm{j}}\frac{2\pi(x_n\sin{\theta_{\ell,k}}\cos{\phi_{\ell,k}}+y_n\cos{\theta_{\ell,k}})}{\lambda}},
\end{align}
\end{subequations}
and where ${\bm\rho}_{\ell,k}=[\sin{\theta_{\ell,k}}\cos{\phi_{\ell,k}},\cos{\theta_{\ell,k}}]^{{T}}$, $L_k$ is the number of propagation paths, $\theta_{\ell,k}$ and $\phi_{\ell,k}$ denote the elevation and azimuth angles of the $\ell$th path, respectively, $\mu_k$ represents the path loss, $\sigma_{k,\ell}$ captures the small-scale fading, and $\lambda$ is the carrier wavelength. To characterize the performance limit, the fusion center is assumed to have perfect CSI.
\subsection{AirComp Model}
At a specific time slot, let $x_k\in{\mathbb{C}}$ denote the parameter collected by node $k$, which satisfies ${\mathbb{E}}\{x_k\}=0$, ${\mathbb{E}}\{\lvert x_k\rvert^2\}=1$, and ${\mathbb{E}}\{x_kx_{k'}\}=0$, $\forall k\ne k'$. Each node $k$ linearly scales its parameter by a power control factor $w_k$ and transmits $w_kx_k$ to the fusion center simultaneously. The fusion center employs a linear detection vector ${\mathbf{u}}=[u_1,\ldots,u_N]^{{T}}\in{\mathbb{C}}^{N\times1}$ to combine the received signals and compute the sum of the transmitted parameters, $x=\sum_{k=1}^{K}x_k$, which is estimated as follows:
\begin{align}\label{MAC_Signal}
\hat{x}={\mathbf{u}}^{{H}}\left(\sum\nolimits_{k=1}^{K}{\mathbf{h}}_kw_kx_k+{\mathbf{n}}\right),
\end{align}
where ${\mathbf{n}}\sim{\mathcal{CN}}({\mathbf{0}},\sigma^2{\mathbf{I}})$ denotes the thermal noise at the fusion center with power $\sigma^2$. The computation distortion is measured by the mean-squared error (MSE) between the desired value $x$ and its estimate $\hat{x}$, which can be written as follows:
\begin{subequations}
\begin{align}
{\rm{MSE}}&\triangleq{\mathbb{E}}\{\lvert{x}-\hat{x}\rvert^2\}\\
&=\sum\nolimits_{k=1}^{K}\lvert{\mathbf{u}}^{{H}}{\mathbf{h}}_kw_k-1\rvert^2+\sigma^2\lVert{\mathbf{u}}\rVert^2.
\end{align}
\end{subequations}
It is important to note that, unlike in FPA-based AirComp systems, the MSE in MA-based AirComp depends on the physical positions of the MAs.
\subsection{Problem Formulation}
\subsubsection{Dynamic Design}
Using existing channel estimation algorithms for MAs \cite{ma2023compressed}, the elevation and azimuth angles ${\mathcal{A}}\triangleq\{\{\theta_{\ell,k}\}_{\ell=1}^{L_k},\{\phi_{\ell,k}\}_{\ell=1}^{L_k},k\in{\mathcal{K}}\}$, as well as the small-scale fading coefficients of each propagation path ${\mathcal{F}}\triangleq\{\{\sigma_{\ell,k}\}_{\ell=1}^{L_k},k\in{\mathcal{K}}\}$, in \eqref{Channel_Model}, can be accurately estimated. With this available channel knowledge, the receive beamformer ${\mathbf{u}}$, the MA locations ${\mathbf{R}}\triangleq[{\mathbf{r}}_1,\ldots,{\mathbf{r}}_N]\in{\mathbb{C}}^{2\times N}$, and the transmit powers of the users can be jointly optimized to minimize the computation error. The corresponding optimization problem is formulated as follows:
\begin{subequations}\label{MSE_Min_Problem}
\begin{align}
&\min\nolimits_{{\mathbf{R}},{}\mathbf{w},{\mathbf{u}}}~{\rm{MSE}}\\
&~{\rm{s.t.}}~\lvert w_k\rvert^2\leq p_k,\forall k\in{\mathcal{K}},\\
&~\quad~~{\mathbf{r}}_n\in{\mathcal{C}},\forall n\in{\mathcal{N}},\lVert {\mathbf{r}}_n-{\mathbf{r}}_{n'}\rVert\leq D,n\ne n',
\end{align}
\end{subequations}
where $p_k$ denotes the transmit power budget of node $k$, $D$ represents the minimum required spacing between antennas to avoid mutual coupling \cite{ivrlavc2010toward}, and ${\mathbf{w}}\triangleq[w_1,\ldots,w_K]^T\in{\mathbb{C}}^{K\times1}$ is the power control vector. Since this scheme requires \emph{instantaneous CSI} and adaptively updates the beamforming, MA positions, and power allocation in real time, it is referred to as the \emph{dynamic design}.

\begin{figure}[!t]
\centering
\includegraphics[width=0.4\textwidth]{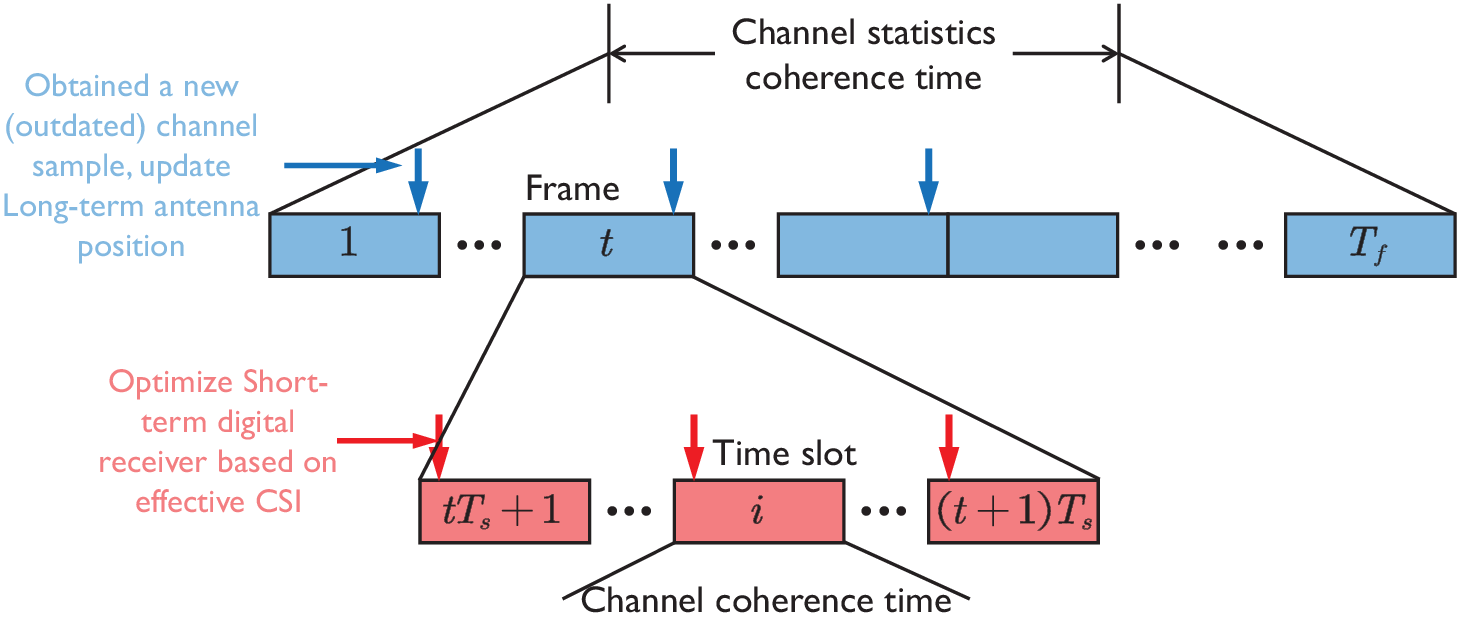}
\caption{Timeline (frame structure) of the static design.}
\label{System_TTS}
\vspace{-20pt}
\end{figure}

\subsubsection{Static Design}
Although instantaneous CSI can be acquired \cite{ma2023compressed}, frequent channel estimation introduces \emph{substantial signaling overhead and resource consumption}, particularly in large-scale systems, thereby degrading overall performance. In multipath environments, the angle information ${\mathcal{A}}$ varies slowly, as it is determined mainly by the geometry of surrounding scatterers. This information remains nearly constant over several coherence intervals and can be obtained from a channel knowledge map. In contrast, the small-scale fading coefficients ${\mathcal{F}}$ vary rapidly across coherence intervals.

Since the design of the power control vector $\mathbf{w}$ and receive beamformer ${\mathbf{u}}$ depends only on the effective channels $\{{\mathbf{h}}_k\}_{k=1}^{K}$, we propose a \emph{static design} that optimizes the MA locations based on long-term channel information ${\mathcal{A}}$ (i.e., angle statistics) and updates $\mathbf{w}$ and ${\mathbf{u}}$ using short-term effective CSI $\{{\mathbf{h}}_k\}_{k=1}^{K}$. This approach significantly reduces signaling overhead while maintaining high performance. As illustrated in {\figurename} {\ref{System_TTS}}, the coherence interval of the small-scale fading statistics is divided into $T_f$ frames, each with $T_s$ time slots. The MA positions are updated at the beginning of each coherence interval, whereas beamforming and power control are adapted per time slot. The corresponding optimization problem is given as follows:
\begin{subequations}\label{AMSE_Min_Problem}
\begin{align}
&\min\nolimits_{{\mathbf{R}}}~{\mathbb{E}}_{\mathcal{F}}\{\min\nolimits_{{\mathbf{w}},{\mathbf{u}}}{\rm{MSE}}\}\\
&~{\rm{s.t.}}~\lvert w_k\rvert^2\leq p_k,\forall k\in{\mathcal{K}},\\
&~\quad~~{\mathbf{r}}_n\in{\mathcal{C}},\forall n\in{\mathcal{N}},\lVert {\mathbf{r}}_n-{\mathbf{r}}_{n'}\rVert\leq D,n\ne n'.
\end{align}
\end{subequations}

It is worth noting that the static design incurs lower channel estimation overhead than the dynamic design but generally results in higher computation error. Both the dynamic design problem in \eqref{MSE_Min_Problem} and the static design problem in \eqref{AMSE_Min_Problem} are challenging to solve due to their nonconvex constraints and the strong coupling among optimization variables. The static design problem is even more difficult, as the expectation of the MSE in \eqref{AMSE_Min_Problem} does not admit a tractable closed-form expression. In the sequel, we develop efficient algorithms to obtain high-quality solutions for both problems.
\section{Dynamic Design}\label{Section: Dynamic Design}
We commence the dynamic design by addressing problem \eqref{MSE_Min_Problem}. To decouple the optimization variables, we adopt the block coordinate descent (BCD) framework by partitioning the variable set into $K+N+1$ disjoint blocks, i.e., $\{\mathbf{u}\}\cup\{w_k\}_{k=1}^{K}\cup\{\mathbf{r}_n\}_{n=1}^{N}$. Accordingly, $K+N+1$ subproblems are solved sequentially, each optimizing one variable block $\mathbf{u}$, $w_k$, or $\mathbf{r}_n$ while keeping the others fixed. By iteratively updating these blocks in an alternating manner, the proposed algorithm converges to a locally optimal solution of problem \eqref{MSE_Min_Problem}.

Given $\{w_k\}_{k=1}^{K}\cup\{\mathbf{r}_n\}_{n=1}^{N}$, the subproblem with respect to the receive beamformer ${\mathbf{u}}$ can be written as follows:
\begin{align}
\min\nolimits_{{\mathbf{u}}}~{\rm{MSE}}=\sum\nolimits_{k=1}^{K}\lvert{\mathbf{u}}^{{H}}{\mathbf{h}}_kw_k-1\rvert^2+\sigma^2\lVert{\mathbf{u}}\rVert^2,
\end{align}
which is a standard convex optimization problem. The optimal solution is obtained by setting the derivative of the MSE with respect to $\mathbf{u}$ to zero, i.e.,
\begin{align}
\frac{\partial{\rm{MSE}}}{\partial{\mathbf{u}}}=\sum_{k=1}^{K}\lvert w_k\rvert^2{\mathbf{h}}_k{\mathbf{h}}_k^{\mathsf{H}}\mathbf{u}+\sigma^2\mathbf{u}-
\sum_{k=1}^{K} w_k\mathbf{h}_k=0.
\end{align}
It follows that the optimal receive beamformer is given in closed form by
\begin{align}\label{Dynamic_Optimal_Receiver}
{\mathbf u}^{\star}=\Big(\sum\nolimits_{k=1}^{K}\lvert w_k\rvert^2{\mathbf{h}}_k{\mathbf{h}}_k^{\mathsf{H}}+\sigma^2\mathbf{I}\Big)^{-1}\Big(\sum\nolimits_{k=1}^{K} w_k\mathbf{h}_k\Big).
\end{align}

Next, given $\{\mathbf{u}\}\cup\{w_{k'}\}_{k'\ne k}\cup\{\mathbf{r}_n\}_{n=1}^{N}$, the subproblem for optimizing $w_k$ is formulated as follows:
\begin{align}\label{Dynamic_Optimal_Power_Control}
{w}_k^{\star}=\argmin\nolimits_{\lvert w_k\rvert^2\leq p_k}\lvert{\mathbf{u}}^{\mathsf{H}}{\mathbf{h}}_kw_k-1\rvert^2. 
\end{align}
which is a standard convex quadratic optimization problem. Its optimal solution is
\begin{align}\label{Dynamic_Optimal_Power}
w_k^{\star}=\frac{{\mathbf{h}}_k^{\mathsf{H}}{\mathbf{u}}}{\lambda+\lvert{\mathbf{u}}^{\mathsf{H}}{\mathbf{h}}_k\rvert^2},
\end{align}
where $\lambda$ denotes the Lagrange multiplier associated with the power constraint. According to the Karush-Kuhn-Tucker (KKT) conditions, the solution satisfies 
\begin{align}
\lambda=\left\{
             \begin{array}{ll}
             0, &  \lvert{\mathbf{u}}^{\mathsf{H}}{\mathbf{h}}_k\rvert^2\geq p_k^{-1}\\
             \frac{\lvert{\mathbf{u}}^{\mathsf{H}}{\mathbf{h}}_k\rvert}{\sqrt{p_k}}-\lvert{\mathbf{u}}^{\mathsf{H}}{\mathbf{h}}_k\rvert^2, & {\rm{Otherwise}}
             \end{array}
\right..
\end{align}

Finally, the marginal subproblem for optimizing the MA position ${\mathbf{r}}_n$ is formulated as follows:
\begin{align}\label{MA_Subproblem}
\max_{{\mathbf{r}}_n\in{\mathcal{S}}_n}f_n\triangleq\sum_{k=1}^{K}(2\Re\{h_k^{*}({\mathbf{r}}_n)c_{k,n}\}-d_{k,n}\lvert h_k({\mathbf{r}}_n)\rvert^2),
\end{align}
where $c_{k,n}=u_nw_k^{*}-\lvert w_k\rvert^2\sum_{n'\ne n}u_nu_{n'}^{*}h_k({\mathbf{r}}_{n'})$, $d_{k,n}=\lvert w_ku_n\rvert^2$, and ${\mathcal{S}}_n\triangleq\{{\mathbf{x}}|{\mathbf{x}}\in{\mathcal{C}},\lVert {\mathbf{x}}-{\mathbf{r}}_{n'}\rVert\leq D,\forall n\ne n'\}$. Since $f_n$ is highly non-convex and lacks a closed-form solution, stationary points of subproblem \eqref{MA_Subproblem} can be obtained using the gradient descent method with backtracking line search \cite{boyd2004convex}. The gradient of $f_n({\mathbf{r}}_n)$ with respect to ${\mathbf{r}}_n$ is computed as follows:
\begin{equation}\label{Der_MA}
\begin{split}
&\nabla_{{\mathbf{r}}_n}f_{n}=\sum_{k=1}^{K}\!\sum_{\ell=1}^{L_i}\!\frac{\lvert\tau_{n}^{k,\ell}\rvert}{-\frac{\lambda}{4\pi}}\sin\!\left(\!\frac{2\pi}{\lambda}
{\mathbf{r}}_n^{{T}}{\bm\rho}_{k,\ell}\!+\!\angle\tau_{n}^{k,\ell}\!\right)\!{\bm\rho}_{k,\ell}\\
&+\!\sum_{k=1}^{K}\!\sum_{\ell=1}^{L_k}\!\sum_{\ell'\ne \ell}\!\frac{\lvert\sigma_{\ell,k}\sigma_{\ell',k}\rvert}{\frac{\lambda}{4\pi d_{k,n}}}\sin\!\left(\!\frac{2\pi}{\lambda}
{\mathbf{r}}_n^{{T}}{\bm\rho}_k^{\ell,\ell'}\!+\!\theta_k^{\ell,\ell'}\!\right)\!{\bm\rho}_k^{\ell,\ell'},
\end{split}
\end{equation}
where $\tau_{n}^{k,\ell}=\sigma_{k,\ell}^{*}c_{k,n}$, ${\bm\rho}_k^{\ell,\ell'}={\bm\rho}_{k,\ell}-{\bm\rho}_{k,\ell'}$, and $\theta_k^{\ell,\ell'}=\angle{\sigma}_{\ell',k}-\angle{\sigma}_{\ell,k}$. The overall optimization for the MA positions is summarized in Algorithm \ref{Algorithm1}, which is guaranteed to converge to a stationary-point solution \cite{boyd2004convex}. The computational complexity of Algorithm \ref{Algorithm1} scales as ${\mathcal{O}}(IN(\sum_{k=1}^{K}L_k^2)\log_2\frac{1}{u_{\min}})$, where $I$ denotes the number of iterations and $u_{\min}$ represents the target accuracy.

\begin{algorithm}[!t]

  \algsetup{linenosize=\tiny} \scriptsize
  \caption{Gradient-Based Algorithm for Optimizing $\mathbf{R}$}
  \label{Algorithm1}
  \begin{algorithmic}[1]
    \STATE Initialize ${\mathbf{R}}^{0}=[{\mathbf{r}}_1^0 \ldots {\mathbf{r}}_N^0]$, the maximum iteration number $I$, step size $u_{\rm{ini}}$, the minimum tolerance step size $u_{\min}$, and set the current iteration $a=0$;
    \REPEAT
    \FORALL{$n=1:N$} 
    \STATE Compute the gradient $\nabla_{{\mathbf{r}}_n^{a}}{f_n}$ and set $u=u_{\rm{ini}}$;
      \REPEAT
      \STATE Compute $\hat{\mathbf{r}}_n={\mathbf{r}}_n^{a}+u\cdot\nabla_{{\mathbf{r}}_n^{a}}{f_n}$ and set $u=u/2$;     
      \UNTIL{$\hat{\mathbf{r}}_n\in{\mathcal{S}}_n \& f_n(\hat{\mathbf{r}}_n)>f_n({\mathbf{r}}_n^{a})$ or $u<u_{\min}$};
      \STATE Set ${\mathbf{r}}_n^{a}=\hat{\mathbf{r}}_n$ and update ${\mathbf{r}}_n^{a+1}=\hat{\mathbf{r}}_n$;
    \ENDFOR
      \STATE Update $a=a+1$;
    \UNTIL{convergence or the maximum iteration number $I$ is reached}.
  \end{algorithmic}
\end{algorithm} 

\begin{algorithm}[!t]
  \algsetup{linenosize=\tiny} \scriptsize
  \caption{BCD-Based Algorithm for Solving Problem \eqref{MSE_Min_Problem}}
  \label{Algorithm2}
  \begin{algorithmic}[1]
    \STATE Initialize $\{{\mathbf{u}}^{0},{\mathbf{w}}^{0},{\mathbf{R}}^{0}\}$, the maximum iteration number $I_{\rm{bcd}}$, and set the current iteration $t=0$;
    \REPEAT
      \STATE Update ${\mathbf{u}}^{t+1}\leftarrow{\mathbf u}^{\star}$;
      \STATE Update ${\mathbf{w}}^{t+1}\leftarrow[w_1^{\star},\ldots,w_{K}^{\star}]^T$;
      \STATE Update ${\mathbf{R}}^{t+1}$ for given $\{{\mathbf{u}}^{t+1},{\mathbf{w}}^{t+1}\}$ by applying the gradient decent method summarized in Algorithm \ref{Algorithm1};
      \STATE Update $t=t+1$;
    \UNTIL{convergence or the maximum iteration number $I_{\rm{bcd}}$ is reached}.
  \end{algorithmic}
\end{algorithm} 

The proposed BCD-based algorithm is summarized in Algorithm \ref{Algorithm2}. We next provide a brief proof of its convergence. For clarity, let $\{{\rm{MSE}}({\mathbf{u}}^{t},{\mathbf{w}}^{t},{\mathbf{R}^{t}})\}$ denote the sequence of objective values in iteration $t$. Since each block update in the BCD framework does not increase the objective value, we have
\begin{equation}\label{Convergence_FP_Proof}
\begin{split}
{\rm{MSE}}({\mathbf{u}}^{t},{\mathbf{w}}^{t},{\mathbf{R}^{t}})\geq{\rm{MSE}}({\mathbf{u}}^{t+1},{\mathbf{w}}^{t+1},{\mathbf{R}^{t+1}}).
\end{split}
\end{equation}
Since ${\rm{MSE}}({\mathbf{u}}^{t},{\mathbf{w}}^{t},{\mathbf{R}^{t}})$ is nonnegative, the sequence $\{{\rm{MSE}}({\mathbf{u}}^{t},{\mathbf{w}}^{t},{\mathbf{R}^{t}})\}$ is monotonically decreasing and bounded below, which guarantees the convergence of Algorithm \ref{Algorithm2}. The computational complexity of the proposed algorithm can be characterized as follows. Let $I_{\rm{bcd}}$ denote the total number of outer BCD iterations. The per-iteration complexity is dominated by the updates of ${\mathbf{u}}$, ${\mathbf{w}}$ and ${\mathbf{R}}$, which respectively scale as $\mathcal{O}(N^3)$, $\mathcal{O}(KN)$, and ${\mathcal{O}}(IN(\sum_{k=1}^{K}L_k^2)\log_2\frac{1}{u_{\min}})$. Therefore, the overall computational complexity of Algorithm \ref{Algorithm2} scales as $\mathcal{O}(I_{\rm{bcd}}(N^3+KN+IN(\sum_{k=1}^{K}L_k^2)\log_2\frac{1}{u_{\min}}))$, which is of polynomial order with respect to the problem dimensions.
\section{Static Design}
We now address the static design. To overcome the mathematical intractability of the MSE expectation in \eqref{AMSE_Min_Problem}, we adopt a stochastic optimization framework inspired by \cite{liu2018online} and develop a stochastic successive convex approximation (SSCA) algorithm. In this approach, the MA locations, i.e., the \emph{long-term} variables, are updated by solving the outer MSE minimization problem in \eqref{AMSE_Min_Problem} using randomly generated channel samples. Meanwhile, the \emph{short-term} variables, including the transmit power control factors at the nodes and the digital receive beamformers at the fusion center, are optimized in each time slot via the BCD-based method presented earlier.
\subsection{Short-Term Optimization Problem}
At each time slot $m\in[tT_s+1,(t+1)T_s]$, the fusion center first estimates the effective fading channel ${\mathbf{H}}(m)\triangleq [{\mathbf{h}}_1(m),\ldots, {\mathbf{h}}_K(m)]$, given the fixed MA positions ${\mathbf{R}}$. It then optimizes the short-term variables, i.e., the power control vector ${\mathbf{w}}$ and the receive beamformer ${\mathbf{u}}$, by solving the following problem:
\begin{align}\label{MSE_Min_SubProblem}
\min\nolimits_{\mathbf{w},{\mathbf{u}}}~{\rm{MSE}}\quad{\rm{s.t.}}~\lvert w_k\rvert^2\leq p_k,\forall k\in{\mathcal{K}},
\end{align}
for the given channel realization ${\mathbf{H}}(m)$. This problem can be efficiently solved using the BCD-based procedure described in Section \ref{Section: Dynamic Design}, where the updates for ${\mathbf{u}}$ and $w_k$ are obtained from \eqref{Dynamic_Optimal_Receiver} and \eqref{Dynamic_Optimal_Power_Control}, respectively. It is important to note that ${\mathbf{w}}$ and ${\mathbf{u}}$ are optimized solely based on the instantaneous effective fading channels $\{{\mathbf{h}}_k\}_{k=1}^{K}$.
\subsection{Long-Term Optimization Problem}
When the statistical CSI of ${\mathcal{F}}$ becomes available, the fusion center optimizes the MA locations ${\mathbf{R}}$ by solving problem \eqref{AMSE_Min_Problem}. However, obtaining a closed-form expression for the optimized MSE after short-term optimization, i.e., $\min\nolimits_{{\mathbf{w}},{\mathbf{u}}}{\rm{MSE}}$, as a function of $\mathbf{R}$ is analytically intractable. To address this challenge, we develop an efficient SSCA algorithm, in which ${\mathbf{R}}$ is iteratively updated by minimizing a convex surrogate function that approximates $\min\nolimits_{{\mathbf{w}},{\mathbf{u}}}{\rm{MSE}}$. 

Specifically, in iteration $t$ of the proposed SSCA-based algorithm, $T_H$ new channel samples of the CSI, denoted by $\{{\mathbf{H}}^{t}(i)\}_{i\in\{1,\ldots,T_H\}}\triangleq\{[{\mathbf{h}}_1(i),\ldots, {\mathbf{h}}_K(i)]\}_{i\in\{1,\ldots,T_H\}}^{t}$, are generated according to the distribution of ${\mathcal{F}}$ and the current MA configuration ${\mathbf{R}}_t$. The average MSE is then approximated by the following recursive concave surrogate function \cite{liu2018online}:
\begin{align}
g_t({\mathbf{R}})=(1-\rho_t)g_{t-1}({\mathbf{R}})+\rho_t{\hat{g}}_t({\mathbf{R}}),
\end{align}
with initialization $g_0({\mathbf{R}})=0$. The step-size sequence $\{\rho_t\}$ satisfies $\rho_t\in(0,1]$, $\sum_{t=1}^{\infty}\rho_t=\infty$, and $\sum_{t=1}^{\infty}\rho_t^2<\infty$. The instantaneous surrogate function ${\hat{g}}_t({\mathbf{R}})$ is constructed as a convex quadratic function \cite{liu2018online}:
\begin{align}
{\hat{g}}_t({\mathbf{R}})=\overline{\rm{MSE}}({\mathbf{R}}_t)+{\mathbf{f}}_{t}^T({\mathbf{r}}-{\mathbf{r}}_t)+\tau_{t}\lVert {\mathbf{r}}-{\mathbf{r}}_t\rVert^2,
\end{align}
where ${\mathbf{r}}={\rm{vec}}({\mathbf{R}})=[{\mathbf{r}}_1^T,\ldots,{\mathbf{r}}_N^T]^T\in{\mathbb{R}}^{2N\times1}$, and $\{\tau_t\}$ are positive constants chosen to ensure the uniform strong convexity of the surrogate function. The approximate gradient ${\mathbf{f}}_{t}\in{\mathbb{R}}^{2N\times1}$ and function value $\overline{\rm{MSE}}({\mathbf{R}}_t)$ are computed as follows:
\begin{align}
{\mathbf{f}}_{t}&\triangleq\frac{1}{T_H}\sum_{i=1}^{T_H}\nabla_{{\mathbf{r}}}{\rm{MSE}}({\mathbf{u}}_{t,i},{\mathbf{w}}_{t,i},{\mathbf{R}_{t}}|{\mathbf{H}}^{t}(i)),\\
\overline{\rm{MSE}}({\mathbf{R}}_t)&\triangleq\frac{1}{T_H}\sum_{i=1}^{T_H}{\rm{MSE}}({\mathbf{u}}_{t,i},{\mathbf{w}}_{t,i},{\mathbf{R}_{t}}|{\mathbf{H}}^{t}(i)),
\end{align}
where $\{{\mathbf{u}}_{t,i},{\mathbf{w}}_{t,i}\}$ denote the suboptimal short-term solutions obtained by solving problem \eqref{MSE_Min_SubProblem} for ${\mathbf{H}}^{t}(i)$, ${\rm{MSE}}({\mathbf{u}}_{t,i},{\mathbf{w}}_{t,i},{\mathbf{R}_{t}}|{\mathbf{H}}^{t}(i))$ is the corresponding MSE, and the gradient of the achievable MSE $\nabla_{{\mathbf{r}}}{\rm{MSE}}({\mathbf{u}}_{t,i},{\mathbf{w}}_{t,i},{\mathbf{R}}|{\mathbf{H}}^{t}(i))$ at ${\mathbf{R}}={\mathbf{R}_{t}}$ can be computed directly from \eqref{Der_MA}.

We then update the MA locations by solving the following surrogate problem:
\begin{subequations}\label{SSCA_SubProblem}
\begin{align}
\min\nolimits_{\mathbf{R}}~&g_t({\mathbf{R}})\\
{\rm{s.t.}}~&{\mathbf{r}}_n\in{\mathcal{C}},\forall n\in{\mathcal{N}},\lVert {\mathbf{r}}_n-{\mathbf{r}}_{n'}\rVert\leq D,n\ne n',
\end{align}
\end{subequations}
where the objective function $g_t({\mathbf{R}})$ is convex and given in closed form. Problem \eqref{SSCA_SubProblem} can be efficiently solved using an element-wise BCD approach. Specifically, we partition ${\mathbf{R}}$ into $2N$ scalar blocks, $\{x_n\}_{n=1}^{N}\cup\{y_n\}_{n=1}^{N}$, and iteratively optimize each coordinate while keeping the others fixed. For instance, the subproblem associated with $y_n$ is expressed as follows:
\begin{align}
\min\nolimits_{y_n\in{\mathcal{Y}}_{n}}~&(1-\rho_t)(y_n[{\mathbf{f}}_{t-1}]_{2n}+\tau_{t-1}(y_n-[{\mathbf{r}}_{t-1}]_{2n})^2)\nonumber\\
&\rho_t(y_n[{\mathbf{f}}_{t}]_{2n}+\tau_{t}(y_n-[{\mathbf{r}}_{t}]_{2n})^2),
\end{align}
where the feasible set is defined as ${\mathcal{Y}}_{n}\triangleq\{y|(x_{n'}-x_{n})^2+(y-y_{n'})^2\geq D^2,n'\ne n, y\in[-A/2,A/2]\}$. When optimizing a single block, the objective function in \eqref{SSCA_SubProblem} reduces to a standard one-dimensional quadratic function, which admits a closed-form optimal solution within the feasible interval. Let $\hat{{\mathbf{R}}}_{t}$ denote the optimized MA position matrix after this step. The long-term variable update is then performed as follows:
\begin{align}
{{\mathbf{R}}}_{t+1}=(1-\gamma_t){{\mathbf{R}}}_{t}+\gamma_t\hat{{\mathbf{R}}}_{t},
\end{align}
where the step-size sequence $\{\gamma_t\}$ satisfies $\gamma_t\in(0,1]$, $\gamma_t\rightarrow0$, $\sum_{t=1}^{\infty}\gamma_t=\infty$, $\sum_{t=1}^{\infty}\gamma_t^2<\infty$, and $\lim_{t\rightarrow\infty}\frac{\gamma_t}{\rho_t}=0$.

\begin{algorithm}[!t]
  \algsetup{linenosize=\tiny} \scriptsize
  \caption{SSCA-Based Algorithm for Solving Problem \eqref{AMSE_Min_Problem}}
  \label{Algorithm3}
  \begin{algorithmic}[1]
    \STATE Initialize ${\mathbf{R}}_{0}$, $g_0({\mathbf{R}})=0$, the maximum iteration number $I_{\rm{ssca}}$, step-size sequences $\{\rho_t\}$ and $\{\gamma_t\}$, number of channel samples $T_H$, and set the current iteration $t=1$;
    \REPEAT
      \STATE Obtain a mini-batch $\{{\mathbf{H}}^{t}(i)\}_{i\in\{1,\ldots,T_H\}}$ based on ${\mathcal{F}}$ and ${\mathbf{R}}_t$;
      \STATE For each ${\mathbf{H}}^{t}(i)$, compute $\{{\mathbf{u}}^{t,i},{\mathbf{w}}^{t,i}\}$ using the BCD-based approach;
      \STATE Construct the surrogate function $g_t({\mathbf{R}})=(1-\rho_t)g_{t-1}({\mathbf{R}})+\rho_t{\hat{g}}_t({\mathbf{R}})$;
      \FORALL{$n=1:N$} 
    \STATE Update $x_n$;
    \STATE Update $y_n$;
    \ENDFOR
    \STATE Update ${{\mathbf{R}}}_{t+1}=(1-\gamma_t){{\mathbf{R}}}_{t}+\gamma_t\hat{{\mathbf{R}}}_{t}$;
      \STATE Update $t=t+1$;
    \UNTIL{convergence or the maximum iteration number $I_{\rm{ssca}}$ is reached}.
  \end{algorithmic}
\end{algorithm}

The proposed SSCA-based algorithm for solving problem \eqref{AMSE_Min_Problem} is summarized in Algorithm \ref{Algorithm3}, where $I_{\rm{ssca}}$ denotes the total number of iterations. The convergence of Algorithm \ref{Algorithm3} is guaranteed under standard assumptions for stochastic approximation \cite{liu2018online}. Specifically, as $t\rightarrow\infty$, the surrogate function ${\hat{g}}_t({\mathbf{R}})$ asymptotically approaches the true objective ${\mathbb{E}}_{\mathcal{F}}\{\min\nolimits_{{\mathbf{w}},{\mathbf{u}}}{\rm{MSE}}\}$, and the generated sequence $\{{\mathbf{R}}_t\}$ converges almost surely to a stationary point of problem \eqref{AMSE_Min_Problem}. The computational complexity of Algorithm \ref{Algorithm3} can be analyzed as follows. First, obtaining the short-term suboptimal solutions $\{{\mathbf{u}}_{t,i},{\mathbf{w}}_{t,i}\}$ for problem \eqref{MSE_Min_SubProblem} requires a complexity of $\mathcal{O}(I_{\rm{bcd}}(N^3+KN))$. Second, the update of ${\mathbf{R}}$ at each iteration using the element-wise BCD method incurs a complexity of ${\mathcal{O}}(N^2)$. Hence, the overall complexity of Algorithm \ref{Algorithm3} scales as ${\mathcal{O}}(I_{\rm{ssca}}(I_{\rm{bcd}}(N^3+KN)+N^2))$.
\section{Numerical Results}\label{Simulations}
In this section, numerical results are presented to validate the performance of the proposed MA-empowered AirComp system. Unless otherwise stated, the key simulation parameters are set as follows: carrier wavelength $\lambda=0.06$ m, number of IoT nodes $K = 6$, transmit power budget $p_k=p=10$ dBm, path loss $\mu_k=-100$ dB, and noise power $\sigma_k^2=-100$ dBm. Each channel consists of $L_k=4$ propagation paths, with small-scale fading coefficients independently drawn from $\sigma_{k,\ell}\sim{\mathcal{CN}}(0,L_k^{-1})$ ($\forall k,\ell$). The minimum antenna separation is set to $D = \frac{\lambda}{2}$, while the elevation and azimuth angles are randomly distributed within $[0,\pi]$. To benchmark performance, we compare the proposed MA-assisted AirComp scheme against a conventional FPA-based system, where the fusion center is equipped with a uniform linear array (ULA) of $N$ fixed antennas separated by $\frac{\lambda}{2}$. For the dynamic design, the BCD-based algorithm is executed with $I_{\rm{bcd}}=200$, $I=50$, $u_{\rm{ini}}=10$, and $u_{\min}=10^{-4}$. For the static design, the SSCA-based algorithm is configured with $T_H=10$, $I_{\rm{ssca}}=2000$, $\tau_t=0.01$, $\rho_t=t^{-1}$, and $\gamma_t=t^{-0.9}$. All variables are initialized randomly.

\begin{figure}[!t]
\centering
\includegraphics[height=0.2\textwidth]{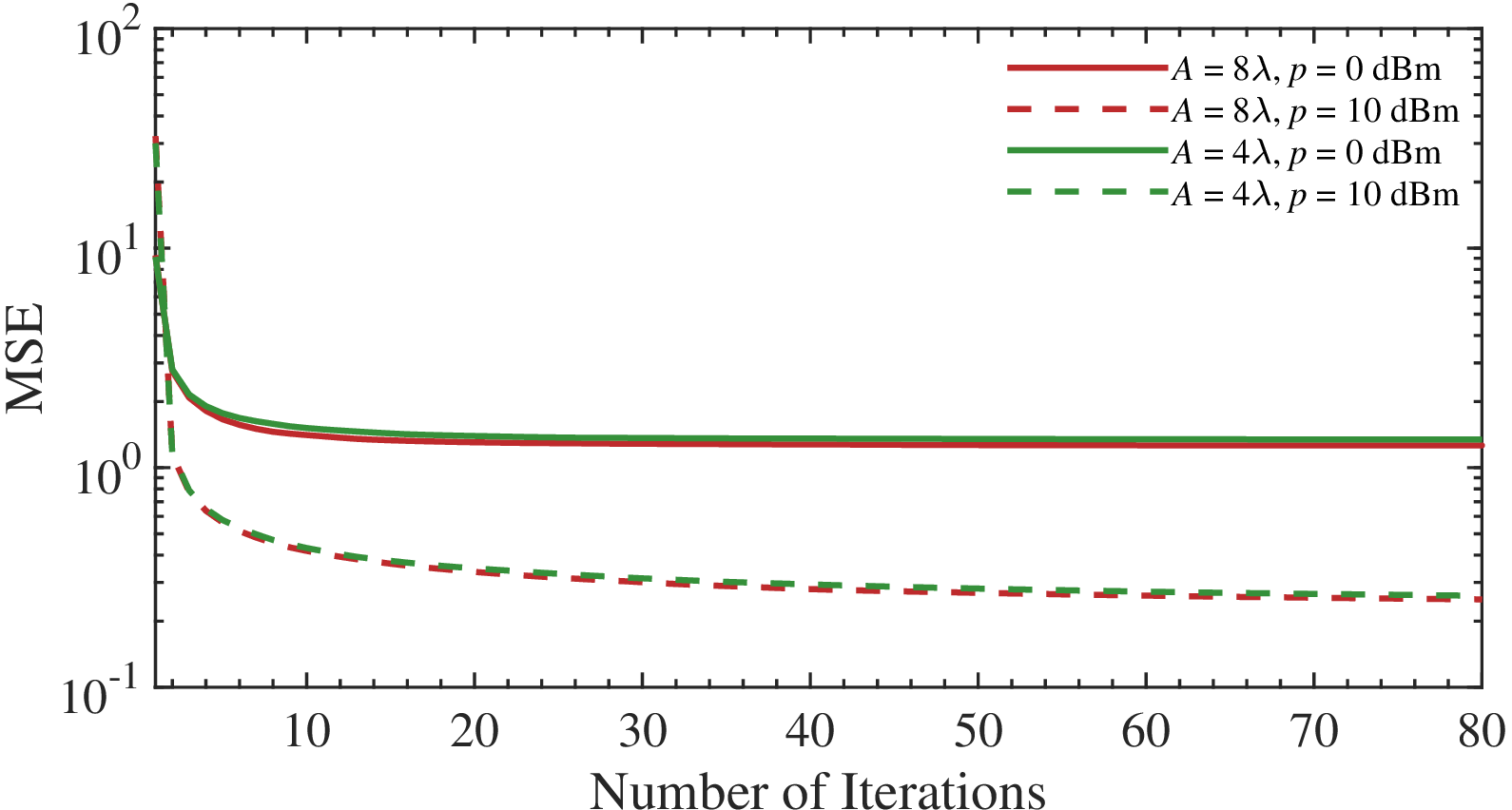}
\caption{Convergence of the proposed dynamic design. $N=6$.}
\label{Figure_MSE_Convergence_Dynamic}
\vspace{-10pt}
\end{figure}

\subsection{Dynamic Design}
We first examine the performance of the MA-enabled AirComp system under the dynamic design. {\figurename} {\ref{Figure_MSE_Convergence_Dynamic}} illustrates the convergence behavior of the proposed BCD-based algorithm. As observed, the MSE decreases monotonically with the number of iterations and converges to a stationary-point solution within approximately $50$ iterations for all considered system settings. Moreover, increasing the transmit power budget of each sensor node significantly improves the computation accuracy, as reflected by a lower MSE. These results confirm the fast convergence and effectiveness of the proposed BCD-based algorithm, thereby validating the theoretical convergence analysis discussed in \eqref{Convergence_FP_Proof}.

\begin{figure}[!t]
\centering
\includegraphics[height=0.2\textwidth]{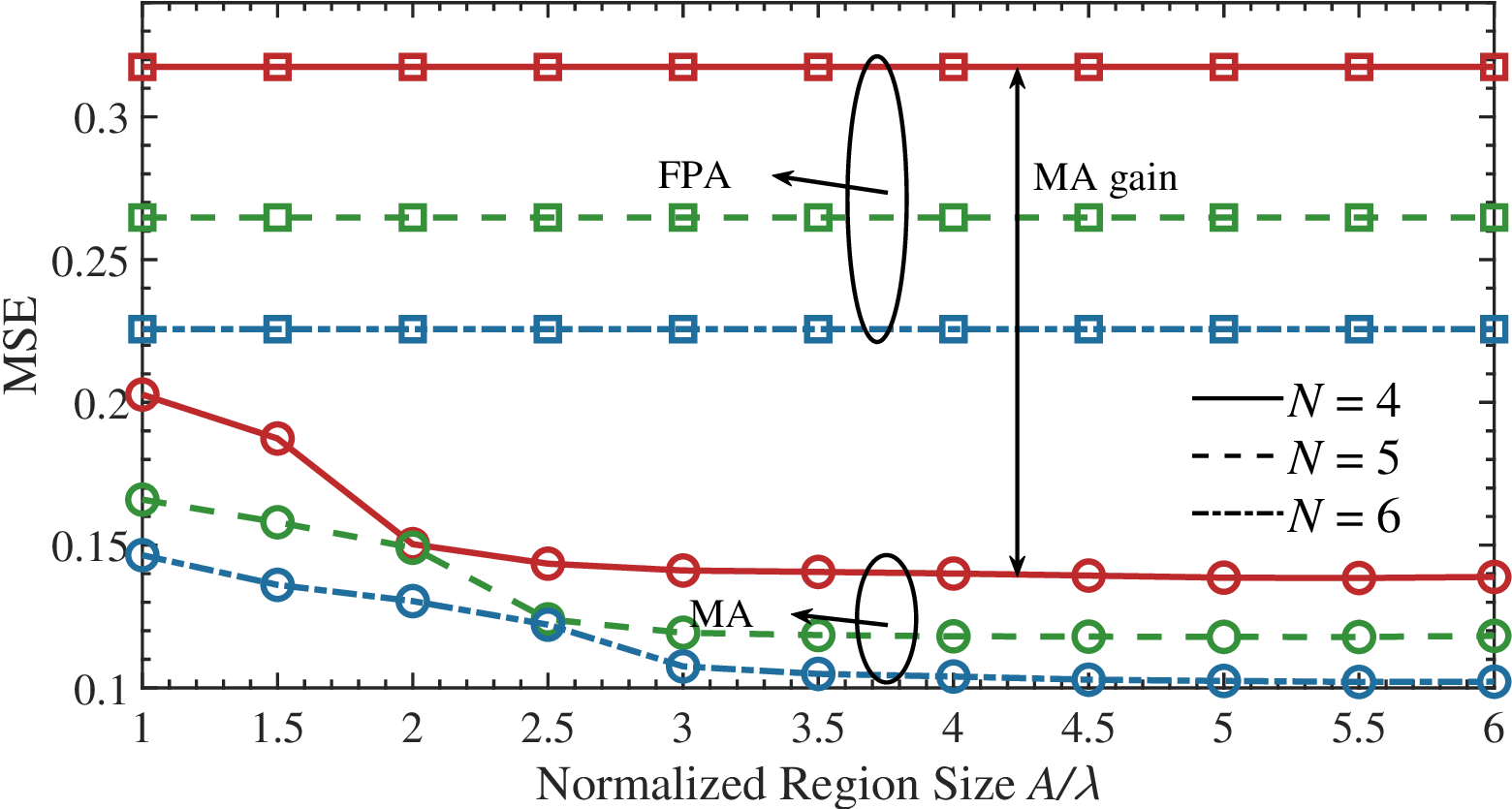}
\caption{MSE versus the aperture size. $K=4$.}
\label{Figure_MSE_Size}
\vspace{-10pt}
\end{figure}

\begin{figure}[!t]
\centering
\includegraphics[height=0.2\textwidth]{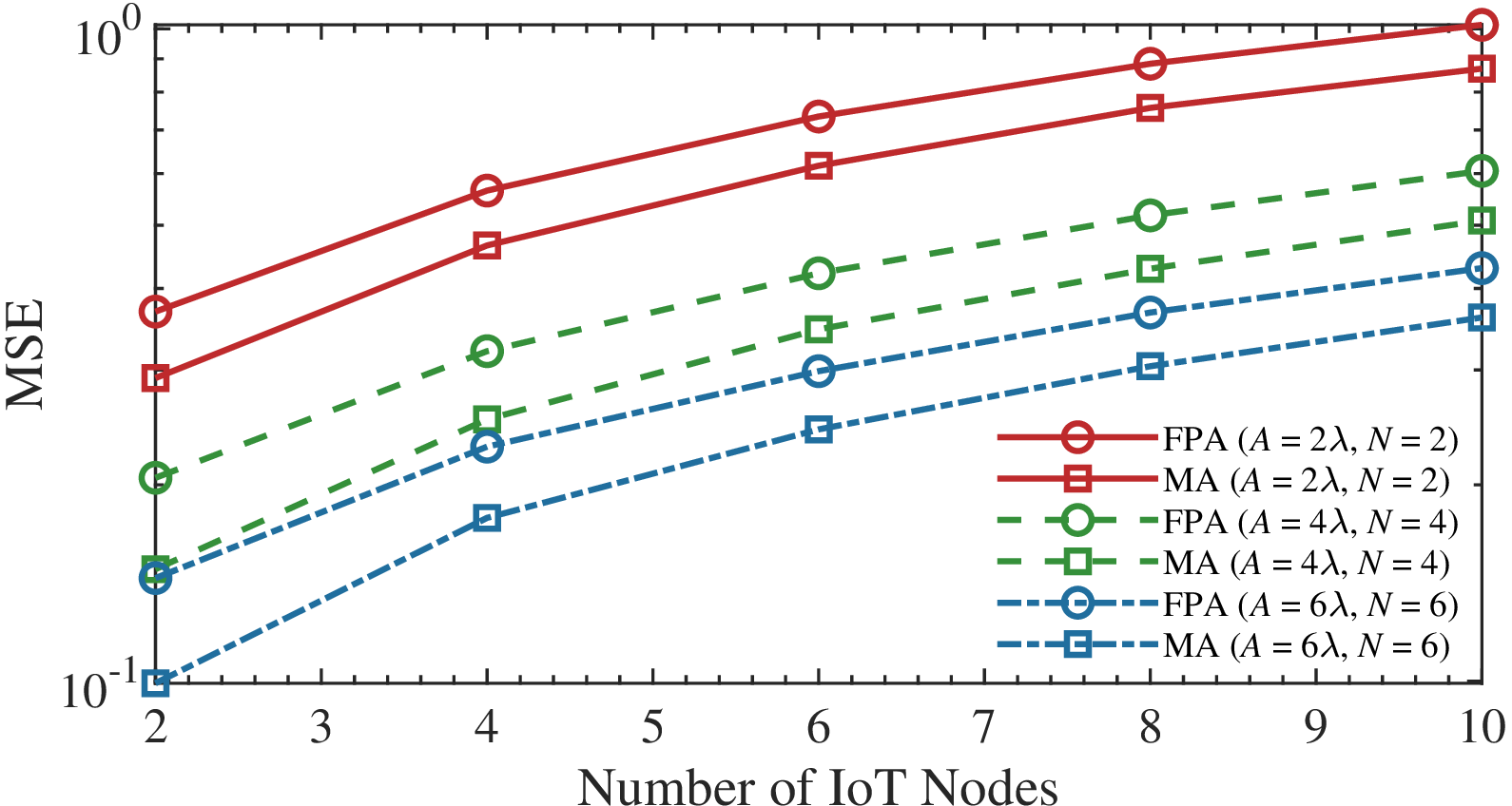}
\caption{MSE versus the number of nodes.}
\label{Figure_MSE_User}
\vspace{-10pt}
\end{figure}

{\figurename} {\ref{Figure_MSE_Size}} compares the MSE performance of the proposed MA-enabled AirComp scheme and the FPA-based benchmark as a function of the aperture size $A$. It can be observed that for a given aperture size, the MA-based design consistently outperforms its FPA counterpart in terms of computation error, thereby validating the effectiveness of antenna location optimization in AirComp systems. The observed MSE improvement primarily stems from the optimized positioning of MAs, which enables spatial adaptation to channel conditions. Moreover, the performance gain of the MA-based system over the FPA-based system becomes more pronounced as the aperture size increases. This is expected since a larger movement region provides additional spatial degrees of freedom for performance enhancement. Interestingly, the MSE achieved by the MA-based system saturates when the normalized aperture size exceeds approximately $A=4\lambda$, indicating that the optimal computation performance can be achieved within a finite spatial region. This observation also explains the negligible MSE improvement when $A$ increases from $4\lambda$ to $8\lambda$, as shown in {\figurename} {\ref{Figure_MSE_Convergence_Dynamic}}.

{\figurename} {\ref{Figure_MSE_User}} further depicts the MSE versus the number of IoT nodes $K$. The results show that the proposed MA-enabled schemes consistently outperform conventional FPA-based systems, and the performance gap widens with increasing aperture size and number of antennas. For all considered cases, increasing $K$ leads to higher computation distortion, as more simultaneous transmissions intensify inter-signal interference during the aggregation process.

\subsection{Static Design}

\begin{figure}[!t]
\centering
\includegraphics[height=0.2\textwidth]{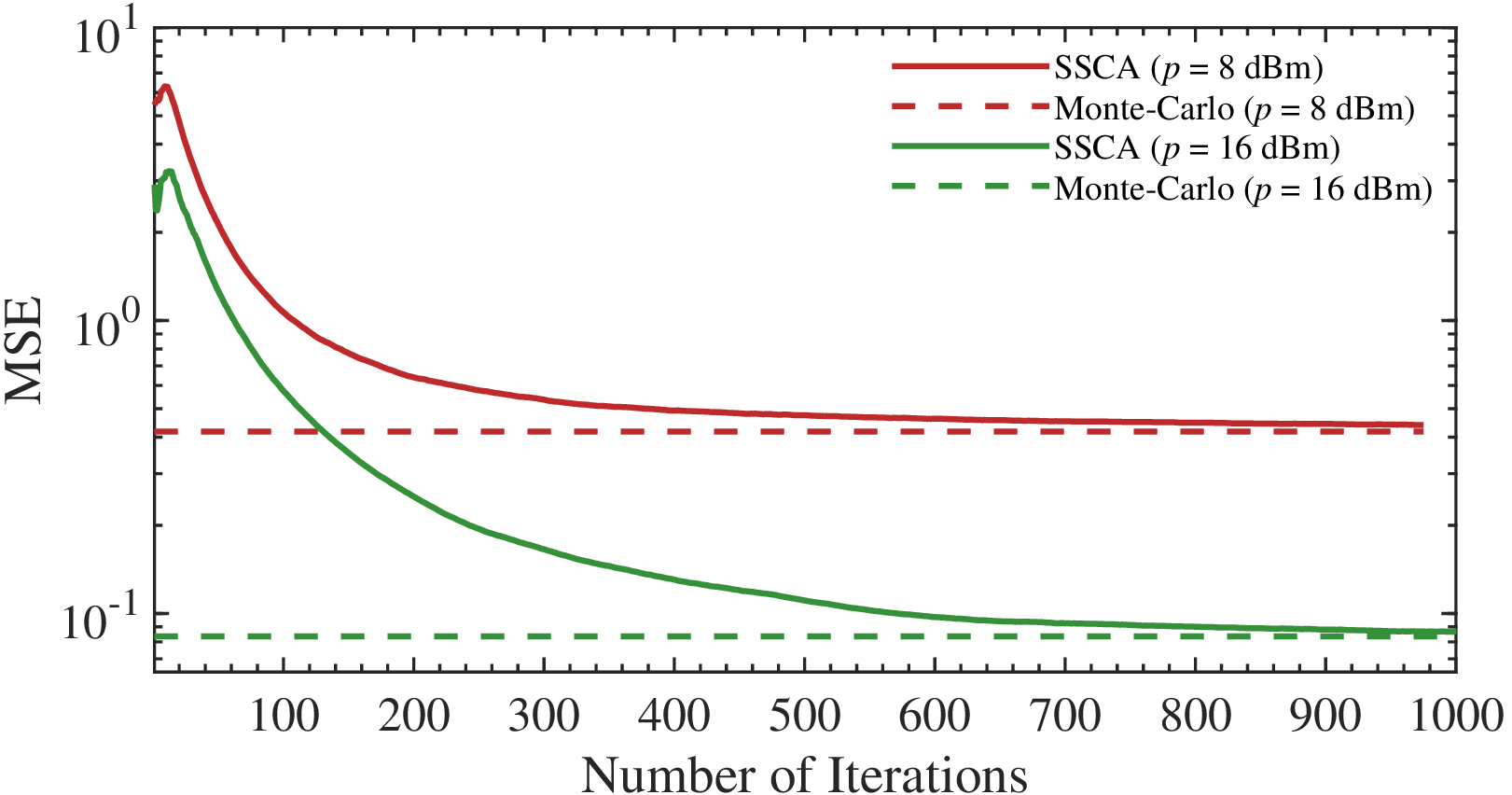}
\caption{Convergence of the proposed dynamic design. $A=8\lambda$ and $N=6$.}
\label{Figure_MSE_Convergence_Static}
\vspace{-10pt}
\end{figure}

\begin{figure}[!t]
\centering
\includegraphics[height=0.2\textwidth]{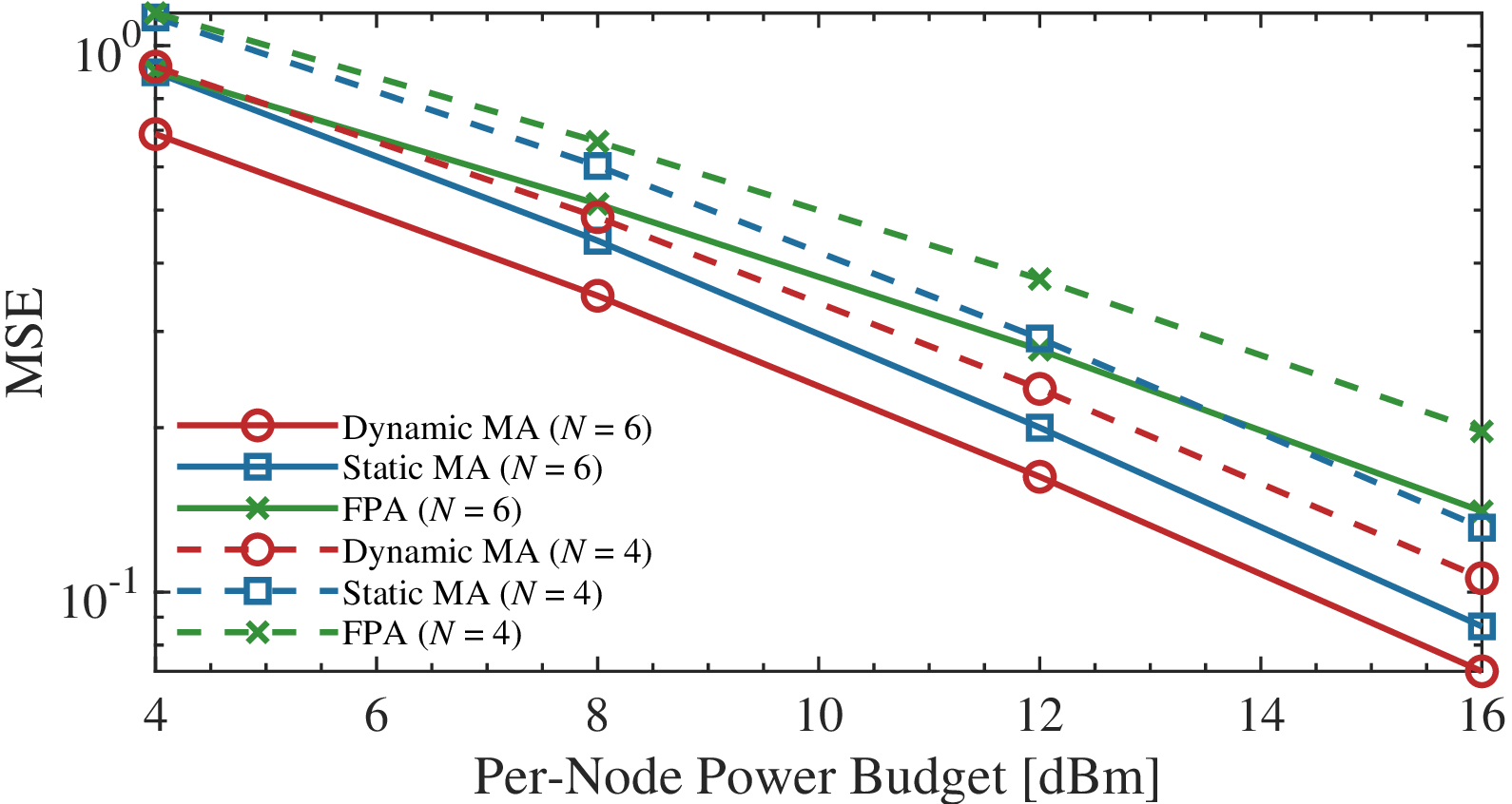}
\caption{MSE comparison of the dynamic design and the static design. $A=8\lambda$.}
\label{Figurw_MSE_Compare}
\vspace{-10pt}
\end{figure}

We next investigate the performance of the static design, where the MA locations are optimized based on long-term channel statistics. {\figurename} {\ref{Figure_MSE_Convergence_Static}} illustrates the convergence behavior of the proposed SSCA-based algorithm. As observed, the MSE achieved by the algorithm decreases steadily with the number of iterations and eventually converges to a stationary-point solution. Moreover, as the iteration index increases, the objective value of the SSCA-based approach gradually converges to the true expectation of the MSE obtained through Monte-Carlo simulations, which validates the theoretical convergence analysis in \cite{liu2018online}. These results confirm the accuracy and convergence stability of the proposed SSCA-based optimization framework.

{\figurename} {\ref{Figurw_MSE_Compare}} compares the performance of the dynamic and static designs under varying transmit power budgets per node. As expected, increasing the power budget improves the computation accuracy for all schemes. The dynamic design consistently outperforms the static one, since it adaptively adjusts the MA positions using real-time channel information and thus fully exploits short-term CSI variations. Nevertheless, both the dynamic and static MA-based designs significantly outperform the conventional FPA-based system, even when only long-term statistical CSI is available. This highlights the effectiveness of employing MAs to enhance AirComp performance. Furthermore, since the static design requires substantially less channel estimation overhead, it is particularly attractive for energy- and resource-efficient AirComp deployments in future large-scale IoT networks.

\section{Conclusion}
In this paper, we proposed a novel MA-empowered AirComp system to reduce computation distortion. Two efficient MSE minimization schemes were developed for the MA-based uplink channel by optimizing the receive antenna positions under both a dynamic design based on short-term CSI and a static design based on long-term CSI. Simulation results validated the superior performance of the proposed MA-enabled architecture compared with existing FPA systems.
\bibliographystyle{IEEEtran}
\bibliography{mybib}
\end{document}